\documentclass[11pt]{article}
\usepackage{amssymb,amsthm,amsmath,float,epsfig}
\usepackage[small,width=0.95\linewidth]{caption}
\usepackage[total={6.5in,9.0in}, top=1.0in, left=1.0in, includefoot]{geometry}
\usepackage{graphicx}
\usepackage{enumerate}
\usepackage{enumitem}
\usepackage{bm}
\usepackage{authblk}
\usepackage{setspace}
\graphicspath{{figures/}}

\newenvironment{se}[1]{\equation\label{eq:#1}\aligned}{\endaligned\endequation}
\newcommand{\bsplit}[1]{\begin{se}{#1}}
\newcommand{\esplit}{\end{se}}

\title{Nonlinear Transport in the Stochastic Standard Map}
\author[1]{Per Sebastian Skardal\thanks{These authors contributed equally to this work}\thanks{Electronic address: \texttt{persebastian.skardals@trincoll.edu}}}
\author[2]{Adam M. Fox$^*$\thanks{Electronic address: \texttt{adam.fox@wne.edu}}}
\affil[1]{Department of Mathematics, Trinity College, Hartford, CT 06106, USA}
\affil[2]{Department of Mathematics, Western New England University, Springfield, MA 01119, USA}

\date{}

\begin{document}

\maketitle

\begin{abstract}
We study a stochastically-driven standard map. The addition of a noise term destroys the invariant manifolds that organize the phase space which allows for more widespread transport than in the noiseless case. Using appropriately defined hitting times to quantify the dynamics, we identify two qualitatively different classes of transport: linear and nonlinear. Linear transport is primarily driven by the stochasticity in the system, while nonlinear transport results from a combination of the nonlinear dynamics and stochasticity and provides a significant speed-up in transport.
\end{abstract}

\section{Introduction}\label{Sec:1}
Symplectic maps represent the discrete-time analogues of Hamiltonian dynamical systems ~\cite{Meiss1992RMP} which have a wide range of applications, for instance in the study of magnetic fields~\cite{Morrison2000PhysPlas} and fluid dynamics~\cite{Aref1984JFM}. Such systems often display rich dynamical phenomena including chaotic trajectories, periodic orbits, and invariant circles, making them a popular subject of interdisciplinary research for mathematicians, physicists, and other scientists~\cite{Aubry1990PhysD}. A particularly well-studied example is the two-dimensional area-preserving map on the cylinder $\mathbb{T}\times\mathbb{R}$ given by
\begin{equation}\label{eq:SM}
f_{0} : \left\{ \begin{array}{l}
x' = x + y' \mod 1,\\
y' = y + \epsilon g(x).
\end{array} \right.
\end{equation}
Here $x$ and $y$ represent angle and action variables, respectively, $x'$ and $y'$ denote their values at the next time iterate, $\epsilon$ is a nonlinearity parameter, and the \emph{force} $g$ is assumed to be smooth and periodic (mod 1). In particular, the choice
\begin{align}
g(x) = \frac{1}{2\pi} \sin(2\pi x) \label{eq:g}
\end{align}
yields Chirikov's standard map~\cite{Chirikov1979PhysRep}, which exhibits a rich ensemble of dynamical phenomena despite its simple form and has therefore been the subject of significant research. 

When $\epsilon=0$ the phase space of the standard map, Eqs.~(\ref{eq:SM}) and (\ref{eq:g}), is foliated by invariant circles with constant action on which the dynamics is a simple rotation with rotation number $\omega = y_0$. KAM Theory guarantees the persistence of invariant circles in the standard map under small perturbation when $\omega$ is ``sufficiently irrational'' \cite{delaLlave01}. For moderate values of $\epsilon$ the dynamics of the standard map can be characterized as either \emph{resonant} or \emph{non resonant}. In the resonant regions, those around where $p \cdot \omega = q$, $(p,q)\in\mathbb{Z}^2$, contractible circles, often referred to as \emph{secondary} circles or \emph{islands}, arise alongside chaotic orbits \cite{Dua08}. These effects are most pronounced near the low-order resonances, or those with small $(p,q)$. The dynamics of the non resonant regions is similar to the dynamics when $\epsilon =0$. Invariant \emph{rotational} circles, homotopic to the circles in the $\epsilon=0$ map, permeate this space, however thin secondary circles and chaotic orbits do exist between them. 

The invariant circles of Eqs.~(\ref{eq:SM})--(\ref{eq:g}), as well as many other systems, organize the phase space and determine the extent of possible transport~\cite{MacKay1984PhysD,Szezech2009Chaos}. Each invariant circle acts as a barrier for motion -- insulating dynamics on either side from one another~\cite{Greene1986PhysD}. Thus, in the noiseless case transport is limited to the movement of trajectories throughout a single invariant set. More widespread transport can only occur once one or more circles acting as barriers are destroyed as a result of an increase in the nonlinearity parameter $\epsilon$~\cite{Bensimon1984PhysD}. Once destroyed, a circle typically gives rise to a {\it cantorus} -- a fractional dimensional cantor set -- that allows a slow ``leaky'' transport~\cite{Baesens1993PhysD,Li1986PRL,MacKay1992NonlinB,Percival1980}. The breakup of these invariant circles in the standard map and other area- and volume-preserving maps has been an active area of research~\cite{Greene1979JMP,Ketoja1989PhysD,MacKay1992NonlinA,MacKay1985CMP,FM14}.

However, all real-world systems -- either natural or man-made -- display some degree of noise or stochasticity, usually in the form of some temporal fluctuations. In many classes of physical systems it has been shown that even very small amounts of noise can fundamentally alter the dynamics of the system~\cite{Bulsara1996PhysTod,Zhou2002PRL}. Investigation into the effect of stochasticity in the context of Eqs.~(\ref{eq:SM}) and other well-studied area- and volume-preserving maps to date have been limited~\cite{Froeschle1975ASS,Karney1982PhysD}. In this paper we study the effects of an added stochastic term in Chirikov's standard map. Specifically, we assume the sinusoidal form of the force $g$, and, after adding a stochastic term $\xi_{\sigma}$, we obtain the \emph{stochastic standard map}
\begin{equation}\label{eq:SSM}
f_{\sigma} : \left\{ \begin{array}{l}
x' = x + y' \mod 1,\\
y' = y + \displaystyle\frac{\epsilon}{2\pi}\sin(2\pi x) + \xi_{\sigma}.
\end{array} \right.
\end{equation}
Here, $\xi_{\sigma}$ is an iid random variable generated at each time iterate. For simplicity, we assume here that $\xi_{\sigma}$ is drawn from the zero-mean normal distribution with variance $\sigma^2$, a tunable parameter that quantifies the noise intensity. We note that the particular shape of the noise distribution does not alter the results we present below, provided that the variance of the noise is $\sigma^2$, but for simplicity we consider Gaussian white noise, i.e., $\xi_\sigma\sim \mathcal{N}(0,\sigma^2)$.

The addition of the stochastic forcing term dramatically alters the dynamics. Most notably, the invariant sets that are so pivotal - fixed points, periodic orbits, and circles - are broken. An immediate consequence is that the action, i.e., the $y$ variable, of all orbits will be unbounded. It is our goal to understand how transport occurs in this new paradigm, and therefore we will study the dynamics in the action (or $y$) direction.

In this paper we employ appropriately defined hitting times to quantify the behavior of the stochastic standard map (\ref{eq:SSM}) and find that the transport is a novel combination of the linear noise and the dynamical nonlinearity. In particular, when the nonlinearity (i.e., the parameter $\epsilon$) is small, we show that the transport is dominated by the noise term and can be well-captured by simple Brownian motion properties. However, for larger nonlinearity we observe that the noise combines with the dynamics to give rise to regions of rapid nonlinear transport. Specifically, there is a significant speed-up in transport in the resonant regions of phase space relative to the non resonant regions. 

The remainder of this paper is organized as follows. In Sec.~\ref{Sec:2} we present a brief survey that demonstrates the effect of temporal noise on the dynamics and define the hitting times we use for quantifying the transport dynamics. In Sec.~\ref{Sec:3} we begin our analysis, first presenting numerical results for hitting times, then presenting an approximation using a simple Brownian motion. In Sec~\ref{Sec:4} we illustrate the scaling properties of the transport and highlight the nonlinear transport effects in the system. In Sec.~\ref{Sec:5} we present an additional demonstration of the nonlinear transport effects. Finally, in Sec.~\ref{Sec:6} we conclude with a discussion of our results.

\section{Survey and Definitions}\label{Sec:2}

\begin{figure*}[ht]
\centering
\epsfig{file =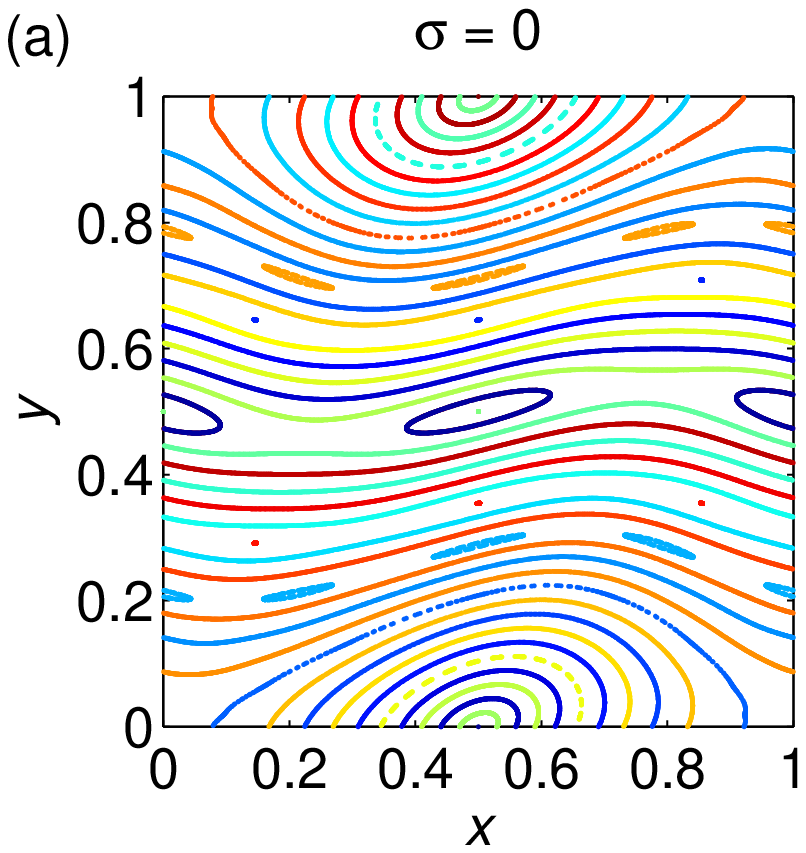, clip =,width=0.32\linewidth }
\epsfig{file =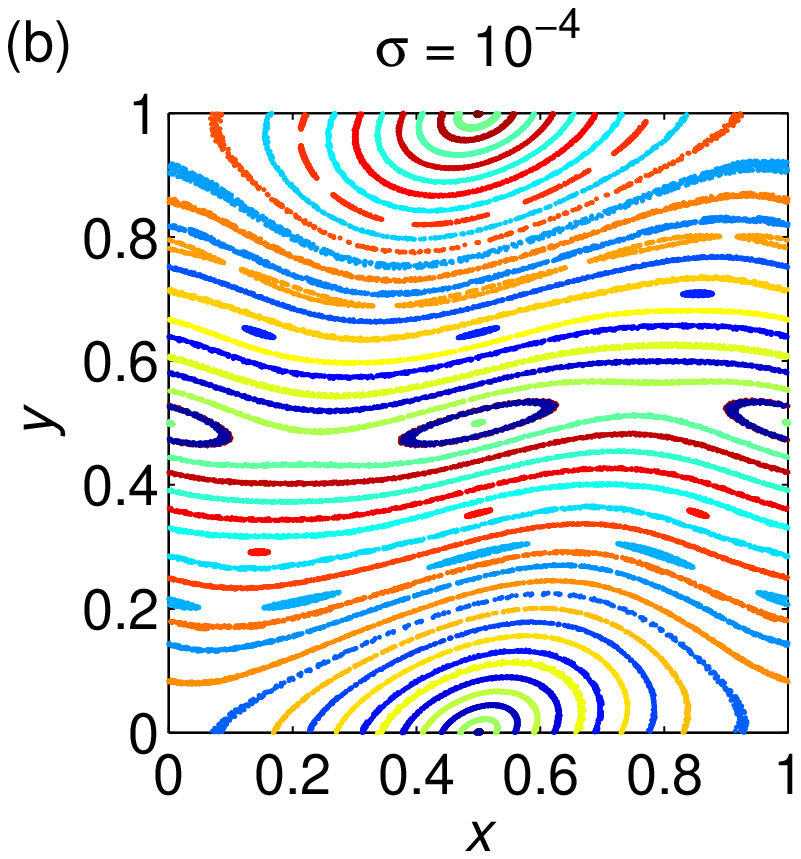, clip =,width=0.32\linewidth } 
\epsfig{file =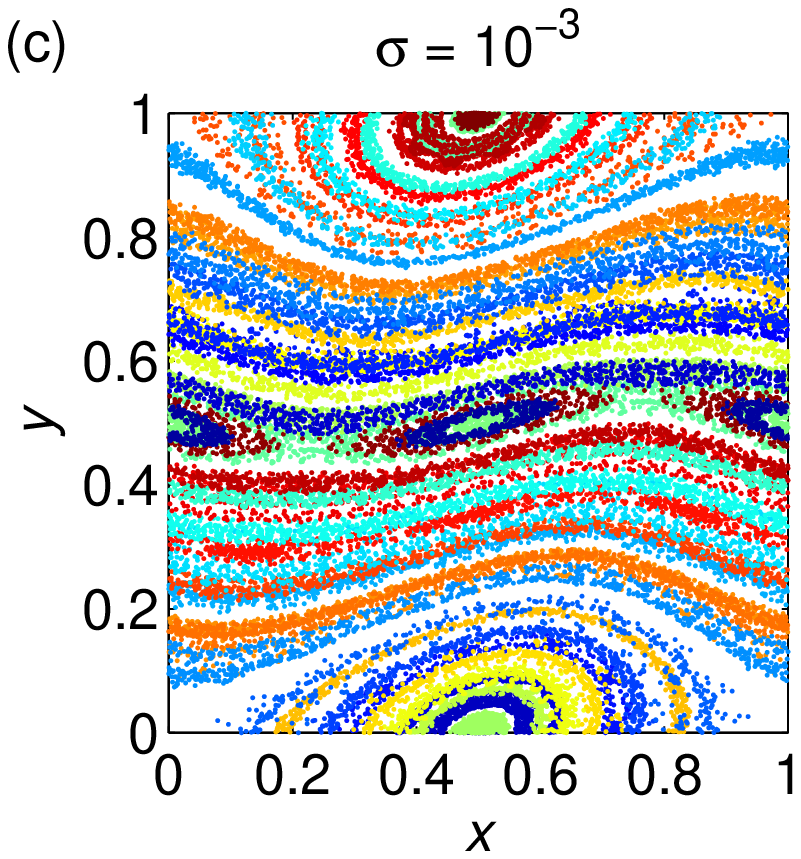, clip =,width=0.32\linewidth } 
\caption{Effect of noise. Phase space of the stochastic standard map (\ref{eq:SSM}) for fixed $\epsilon = 0.5$ and noise intensities $\sigma = 0$ (a), $10^{-4}$ (b), and $10^{-3}$ (c).} \label{fig1}
\end{figure*}

We begin by demonstrating the effect that added stochasticity has on the dynamics of (\ref{eq:SSM}). In particular, we consider three levels of noise: $\sigma=0$ (i.e., no noise), $10^{-4}$, and $10^{-3}$. Setting the nonlinearity parameter to $\epsilon = 0.5$ we simulate (\ref{eq:SSM}), using several different initial conditions for each value of $\sigma$ and plot the results in Fig.~\ref{fig1}. Results for $\sigma=0$, $10^{-4}$, and $10^{-3}$ are shown in panels (a)--(c), respectively, and different colors indicate different orbits obtained using different initial conditions. Each orbit plotted consists of $800$ time iterations. The $\sigma=0$ case [panel (a)] corresponds to the classical standard map, i.e., Eqs.~(\ref{eq:SM})-(\ref{eq:g}) where each orbit shown is bounded away from every other. When $\sigma$ is increased to $10^{-4}$ [panel (b)] we observe a similar structure to the phase space, save for a slight thickening of the orbits. While mixing will eventually occur, the short-term dynamics shown in panel (b) resemble to a remarkable degree those in panel (a). When $\sigma$ is further increased to $10^{-3}$ the thickening of each orbit is made even more pronounced. While the orbits are not as easy to distinguish as in panels (a) or (b), the overall structure of the dynamics is maintained to a certain degree.

This survey clearly illustrates the fundamental change in dynamical behavior that occurs in the standard map with the addition of stochasticity. In particular, the invariance that is key to the dynamics in the noise-less standard map is destroyed. Most importantly, this allows for transport not just within an invariant set in the noise-less case, but across the entire phase space. Evidence for this more robust transport can be first observed in panel (b) ($\sigma=10^{-4}$) with the slight thickening of the orbits. However, transport is more clearly seen in panel (c) ($\sigma=10^{-3}$) as orbits are already overlapping after only $800$ iterations.

To investigate the effect of stochasticity on the new dynamics of (\ref{eq:SSM}) and especially to quantify the transport throughout the phase space, we will use an appropriately defined hitting time. In particular, we examine how long it takes an orbit of the stochastic standard map to reach a distance $a$ from an invariant set of the noiseless standard map. Our exploration of these dynamics will be split into two parts. In Secs~\ref{Sec:3} and \ref{Sec:4} we examine the behavior near a rotational circle of the noiseless system and in Sec~\ref{Sec:5} we study transport near a periodic orbit of the noiseless system. 

More precisely, let
\begin{align}
\Phi_0(x_0,y_0)=\{f_0^t(x_0,y_0)|t\in\mathbb{N}\}.\label{eq:flow02}
\end{align}
denote an orbit of the noise-less map $f_0$ [see Eqs.~(\ref{eq:SM})]. We examine the distance between $\Phi_0(x_0,y_0)$ and orbits of the stochastically forced standard map (\ref{eq:SSM}) with nose level $\sigma$ beginning at the same initial condition $(x_0,y_0)$. Specifically, we calculate for a given distance $a$ the hitting time $\tau_a$, which is defined as the first time $t$ that the orbit of $f_\sigma$ [see Eqs.~(\ref{eq:SSM})] beginning at $(x_0,y_0)$ equals or exceeds a distance $a$ from the set $\Phi_0(x_0,y_0)$:
\begin{align}
\tau_a(x_0,y_0;\sigma)=\inf\{t\in\mathbb{N} \; | \; d[f_\sigma^t(x_0,y_0),\Phi_0(x_0,y_0)]\ge a\}.\label{eq:hittingtime}
\end{align}
Thus, given an appropriate distance metric $d(\cdot,\cdot)$ on the space $\mathbb{T}\times\mathbb{R}$, the hitting time $\tau_a$ represents the time it takes for a noisy trajectory to diverge from the noiseless trajectory by a distance $a$, and here we are interested in the expected hitting time.

\begin{figure*}[ht]
\centering
\epsfig{file =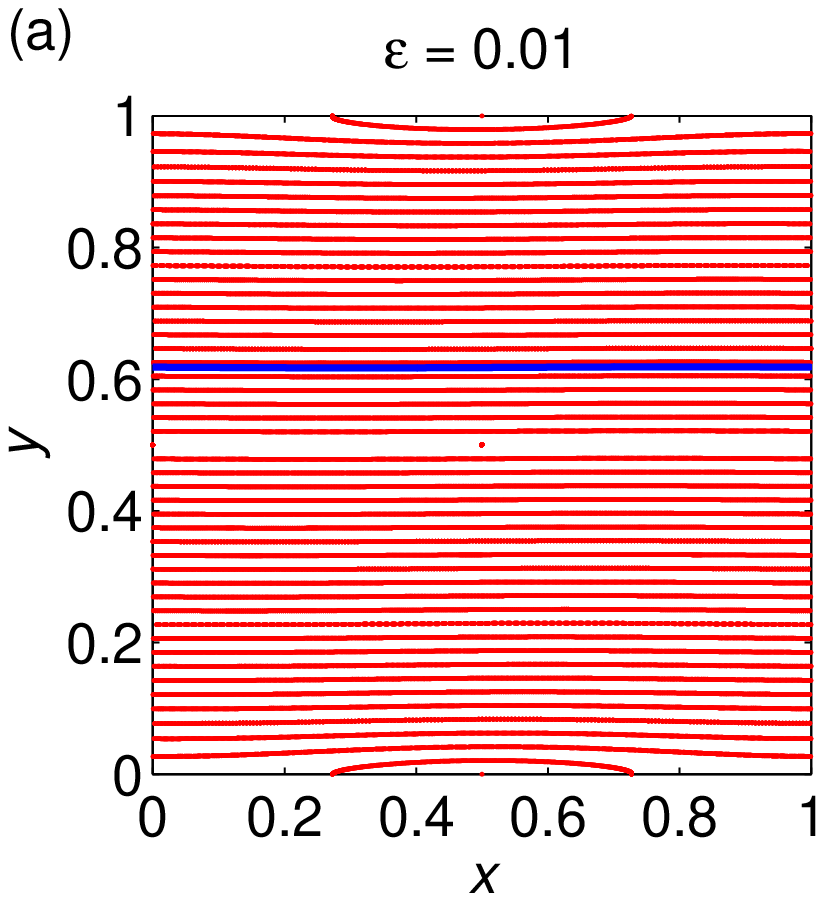, clip =,width=0.32\linewidth }
\epsfig{file =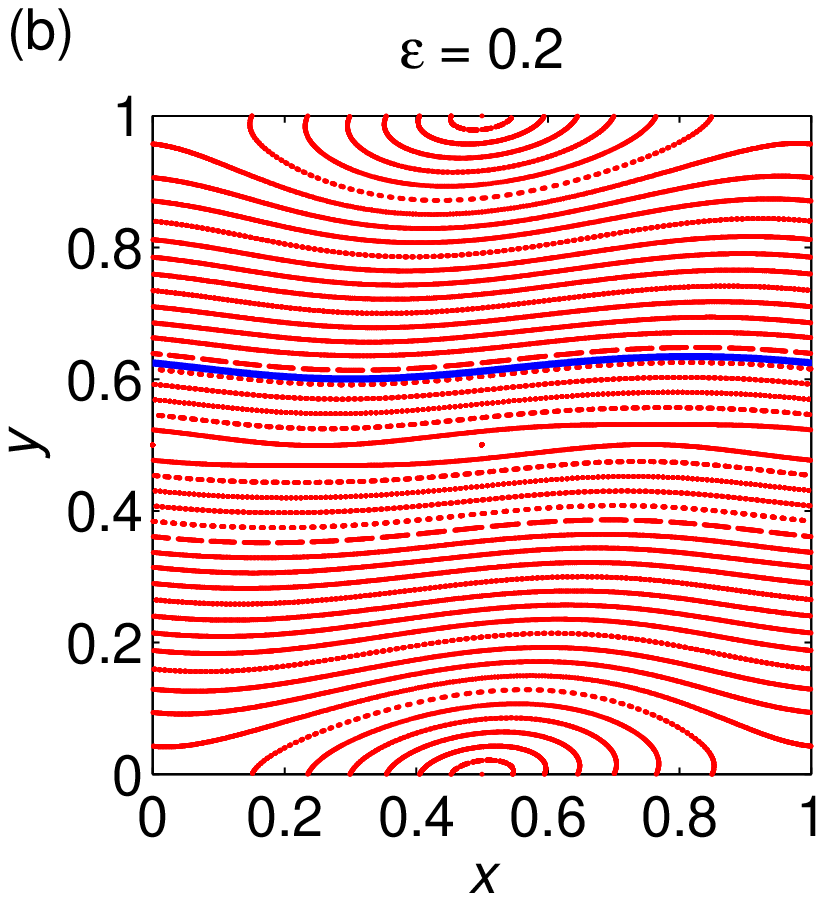, clip =,width=0.32\linewidth }
\epsfig{file =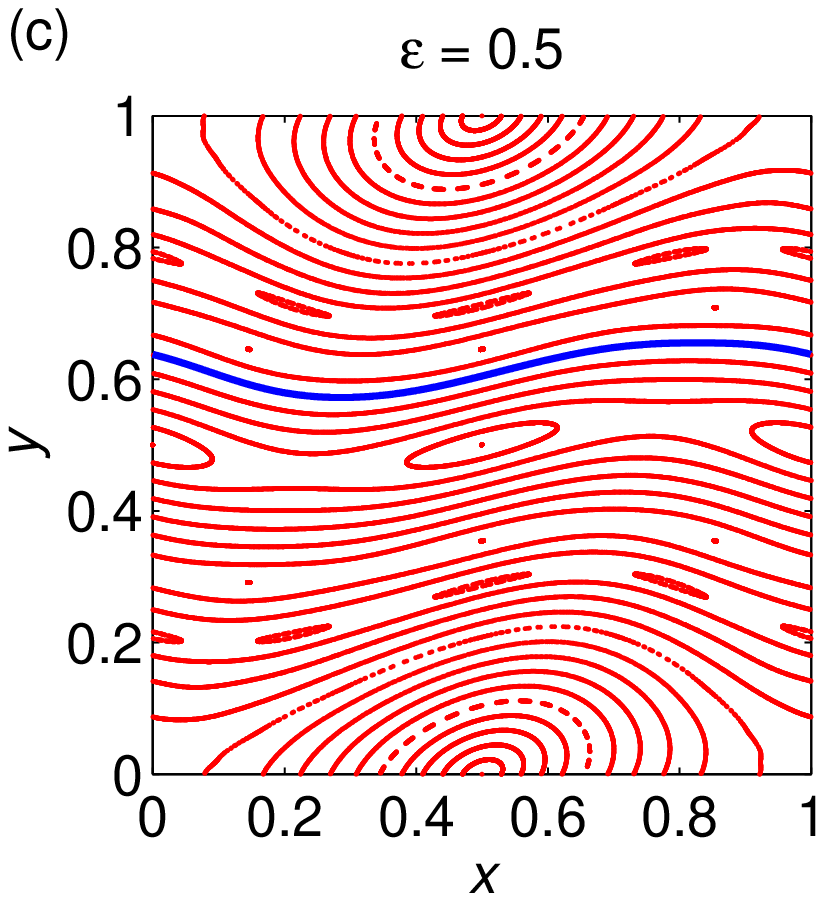, clip =,width=0.32\linewidth }
\caption{Golden-mean circle. For the noise-less standard map, the golden-mean invariant torus for $\epsilon = 0.01$ (a), $0.2$ (b), and $0.5$ (c) as the thick blue curve, compared to other trajectories in red.} \label{fig2}
\end{figure*}

As an illustrative example to begin with, in Sec.~\ref{Sec:3} we draw initial conditions from the {\it golden-mean circle}, the rotational circle with $\omega = \frac{-1+\sqrt{5}}{2}$, which we illustrate in Figs.~\ref{fig2}(a), (b), and (c) for nonlinearity parameters $\epsilon=0.01$, $0.2$, and $0.5$, respectively. The golden-mean circle if denoted as s thick blue curve, and other orbits are denoted in red. We choose the golden-mean circle in particular because of its robustness, i.e., the golden mean circle is believed to be the last circle to survive and is only destroyed when the nonlinearity parameter is increased to $\epsilon=\epsilon_c\approx 0.971635$~\cite{Mac93,Olvera2008SIAM}. Points on this circle are computed using the quasi-Newton method developed by de la Llave et al.~\cite{HdlLS14,FM14}.

Given the topology of the cylinder and the fact that we are most interested in exploring the noise-induced transport in the action variable, we consider the simple distance metric defined solely by the displacement in the $y$-direction. In practice we begin with an initial condition on the circle and iterate $f_0$ 1000 times to generate a sample of points on $\Phi_0$. To measure the distance from a point $(x,y)$ to $\Phi_0$ we use a linear interpolation to approximate the point on the circle $(x_c,y_c)$ with the same $x$ coordinate, $x_c=x$. The distance is then given by $d=|y_c-y|$. This is well-defined since every rotational invariant circle of the standard map is a graph \cite{Meiss1992RMP}. This choice for $d$ is further motivated by a simplification that it will allow in the analysis we present below. 

%In Sec~\ref{Sec:5} every orbit begins at the period-two orbit of $f_0$, $\Phi_0=\left \{ (0,\frac{1}{2}),(\frac{1}{2},\frac{1}{2}) \right \}$. This orbit lies along the $(p,q)=(2,1)$ resonance and is located symmetrically between the largest resonant regions centered at $(\frac{1}{2},0)$ and $(\frac{1}{2},1)$. As we show in Sec~\ref{Sec:5} this symmetry highlights the transition between linear and nonlinear transport. We once again use the displacement in the action to measure distance, i.e. a point $(x,y)$ is said to be a distance $d=|\frac{1}{2}-y|$ from $\Phi_0$. 

\section{Hitting Times and Brownian Motion}\label{Sec:3}

\begin{figure*}[ht]
\centering
\epsfig{file =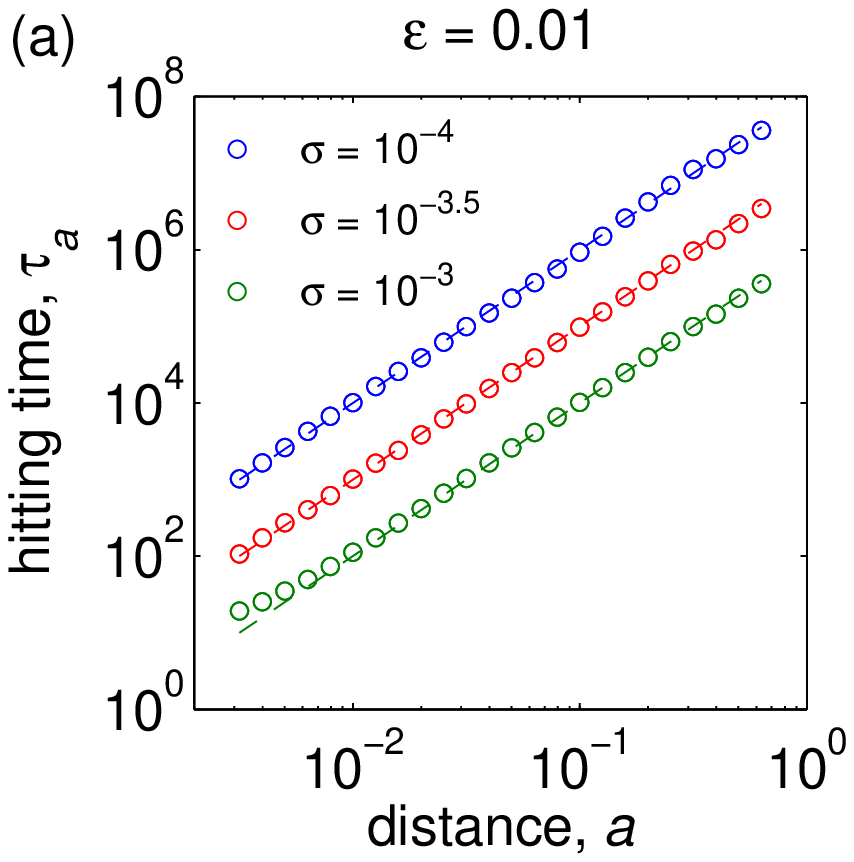, clip =,width=0.32\linewidth }
\epsfig{file =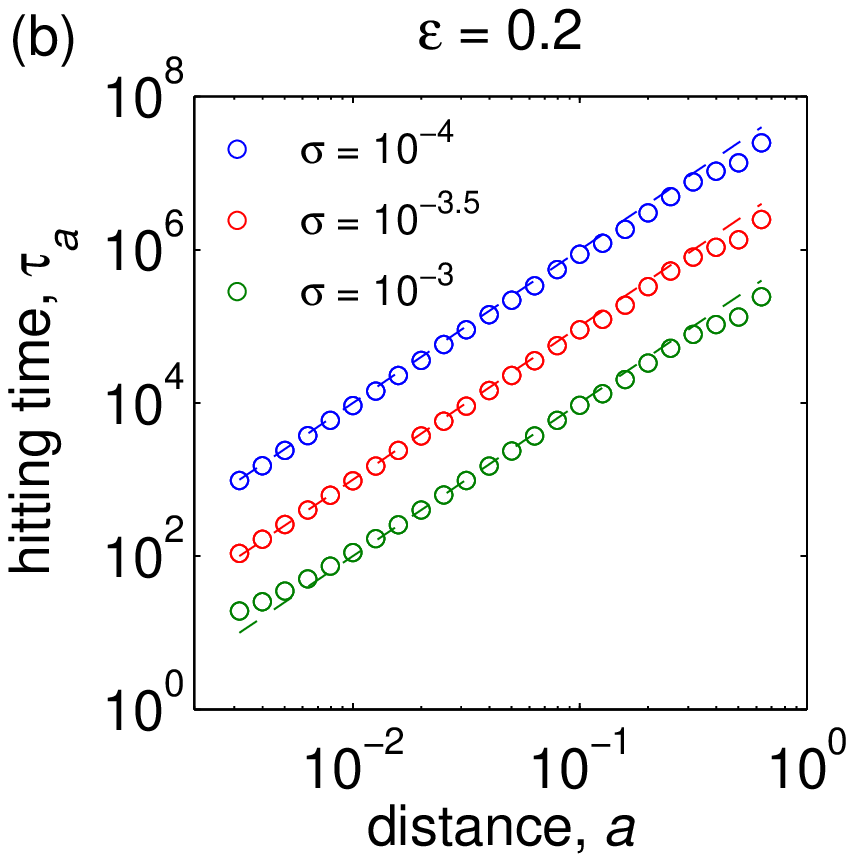, clip =,width=0.32\linewidth } 
\epsfig{file =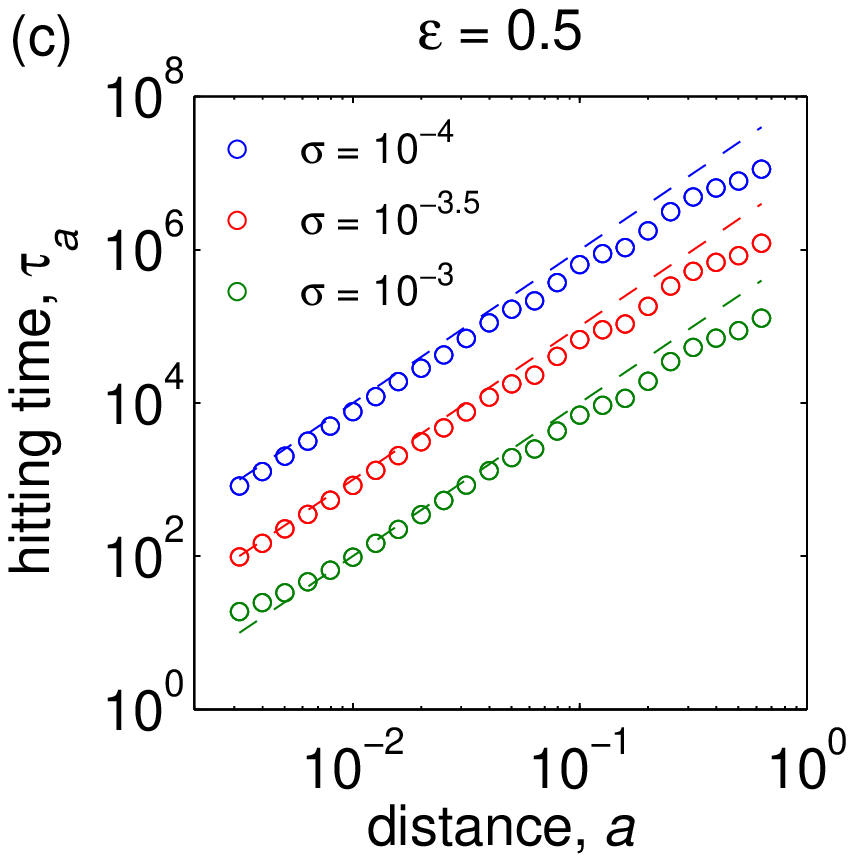, clip =,width=0.32\linewidth } 
\caption{Hitting times. For nonlinearity parameters $\epsilon=0.01$ (a), $0.2$ (b), and $0.5$ (c), we plot the mean hitting time $\tau_a$ vs distance $a$ for noise intensities $\sigma = 10^{-4}$, $10^{-3.5}$, and $10^{-3}$, which are plotting in blue, red, and green circles, respectively. Each data point represents the mean over $10^4$ realizations. The Brownian motion approximations $\tau_a=a^2/\sigma^2$ are plotted as appropriately colored dashed curves.} \label{fig3}
\end{figure*}

Using the hitting times defined above we now study the effect of different noise levels on the dynamics of (\ref{eq:SSM}). Starting on the golden mean circle described above, we calculate the hitting times $\tau_a$ for various distances $a$, nonlinearity parameters $\epsilon$, and noise levels $\sigma$. Since the process is stochastic, we will be interested in the expected hitting times, and therefore we will calculate for each set of parameters the mean $\tau_a$ from $10^4$ realizations. In Fig.~\ref{fig3} we plot the hitting times $\tau_a$ vs the distance $a$ for $\epsilon=0.01$, $0.2$, and $0.5$ in panels (a)--(c), respectively, and for noise levels $\sigma=10^{-4}$, $10^{-3.5}$, and $10^{-3}$, plotted in blue, red, and green circles, respectively. For each value of $\epsilon$, as expected, the expected hitting time $\tau_a$ increases both as the distance $a$ increases and as the noise level decreases. Note, however, that for smaller $\epsilon$, e.g., panel (a), the rate at which $\tau_a$ increases with $a$ is quite regular, increasing very much like a power-law, while for larger $\epsilon$, e.g., panel (c), the rate of increase is much less regular, especially at larger distances $a$.

To gain a better understanding of the behavior of $\tau_a$ we consider the behavior of the process in the limit of small nonlinearity, i.e., $\epsilon\to0$. In particular, we note that in this limit the vertical motion depends only on the noise terms, i.e., $y'-y=\xi_{\sigma}$. Thus, given our notion of distance defined by displacement in the action variable, we ignore motion in the angle variable, obtaining effectively a one-dimensional system. Next, since the noise $\xi_{\sigma}$ is assumed to have variance $\sigma^2$, a simple application of the Central Limit Theorem implies that, at a large enough time $t$, the distribution for the displacement $y_t-y_0$ is Gaussian with variance $\sigma^2t$. It follows that at large enough times the discrete process can thus be approximated as finite-time slices of a one-dimensional Brownian motion with variance $\sigma^2$, for which the hitting time is well known to be given by
\begin{align}
\tau_a=\frac{a^2}{\sigma^2}.\label{eq:Brownian}
\end{align}
Eq.~(\ref{eq:Brownian}) thus provides an approximation for the hitting times for small $\epsilon$, assuming that $\tau_a$ is not too small (i.e., $\sigma$ and $a$ are not simultaneously too big and small, respectively).

To compare this approximation with our results, we plot in Figs.~\ref{fig3}(a)--(c) the Brownian motion approximation $\tau_a=a^2/\sigma^2$ as dashed curves. For the case of $\epsilon=0.01$ [panel (a)], the approximation captures the observed behavior extremely well as the scaling $\tau_a\propto a^2$ is almost exactly obtained. For $\epsilon=0.2$ the approximation remains remarkably accurate for $a$ not too large, until it begins to break down near $a\approx0.4$. Finally, for $\epsilon=0.5$ the approximation breaks down earlier still, near $a\approx0.05$, but provides a good benchmark for smaller $a$. We note that in our simulations there is a small discrepancy for $\sigma=10^{3}$ and small $a$, as the green circles towards the beginning of each plot slightly overshoot the approximation. We find that this is a result of modeling a discrete process with a continuous one - an effect that arises when the expected hitting time is not large enough. These results beg the question of why the approximation fails at certain distances $a$ for larger nonlinearity parameters $\epsilon$ - a point we address next.

\section{Rescaling and Nonlinear Transport}\label{Sec:4}

While the approximation via Brownian motion presented above represents a useful benchmark for understanding the dynamics of the noisy system, two interesting points remain. First, as the Brownian motion approximation fails, we observe a speed-up in the transport of the system. In other words, when the approximation $\tau_a\approx a^2/\sigma^2$ loses accuracy, the observed mean hitting time is always smaller, indicating that transport always occurs quicker - and never slower - than predicted by the Brownian motion approximation. Second, the speed-up in transport of the system seems to scale with the noise level. In particular, for a given value of $\epsilon$, the deviation from the approximation appears the same up to a rescaling for different values of $\sigma$. [See in particular the right-hand-side of Fig.~\ref{fig3}(a), where the blue, red, and green circles appear to undershoot their respective approximations by the same amount in each case.]

\begin{figure*}[ht]
\centering
\epsfig{file =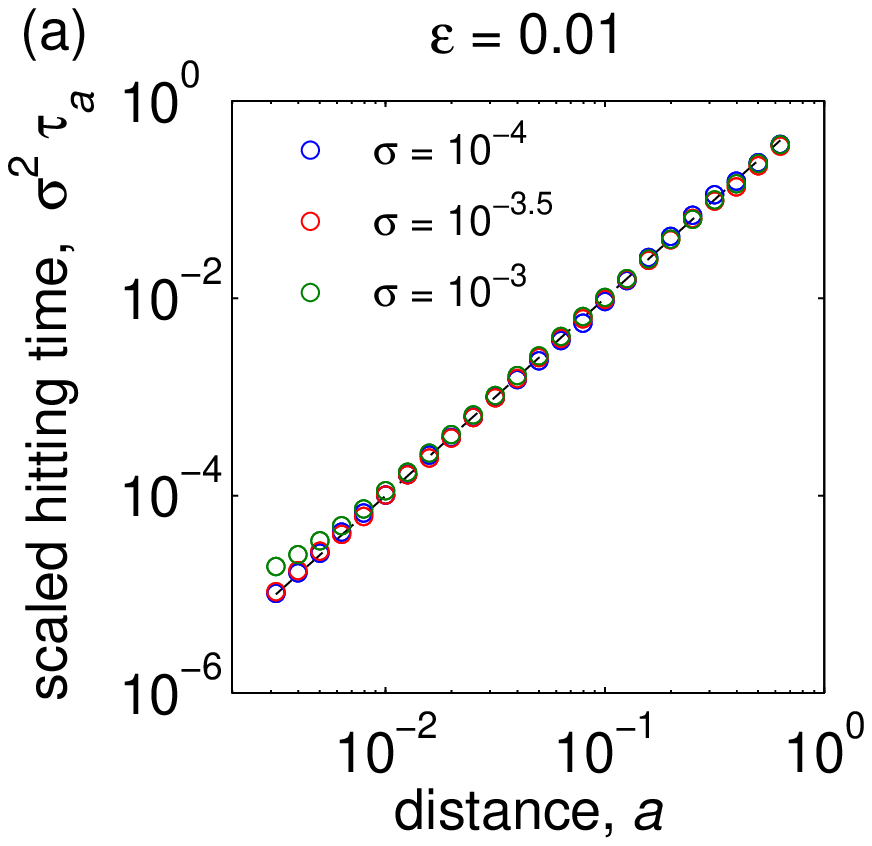, clip =,width=0.32\linewidth }
\epsfig{file =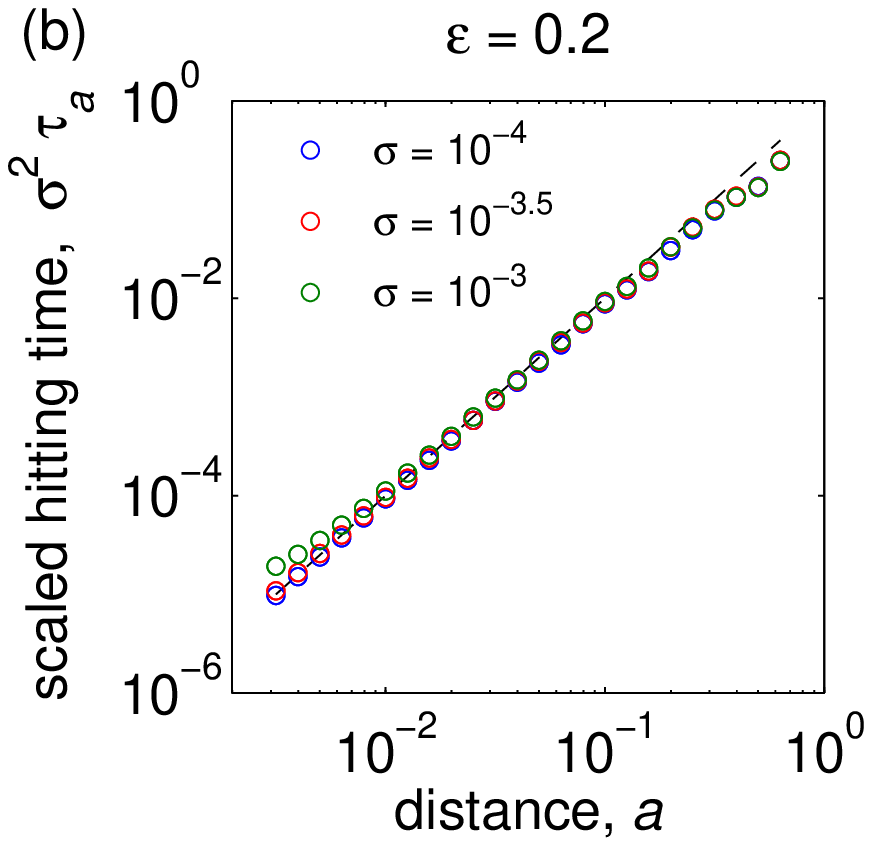, clip =,width=0.32\linewidth } 
\epsfig{file =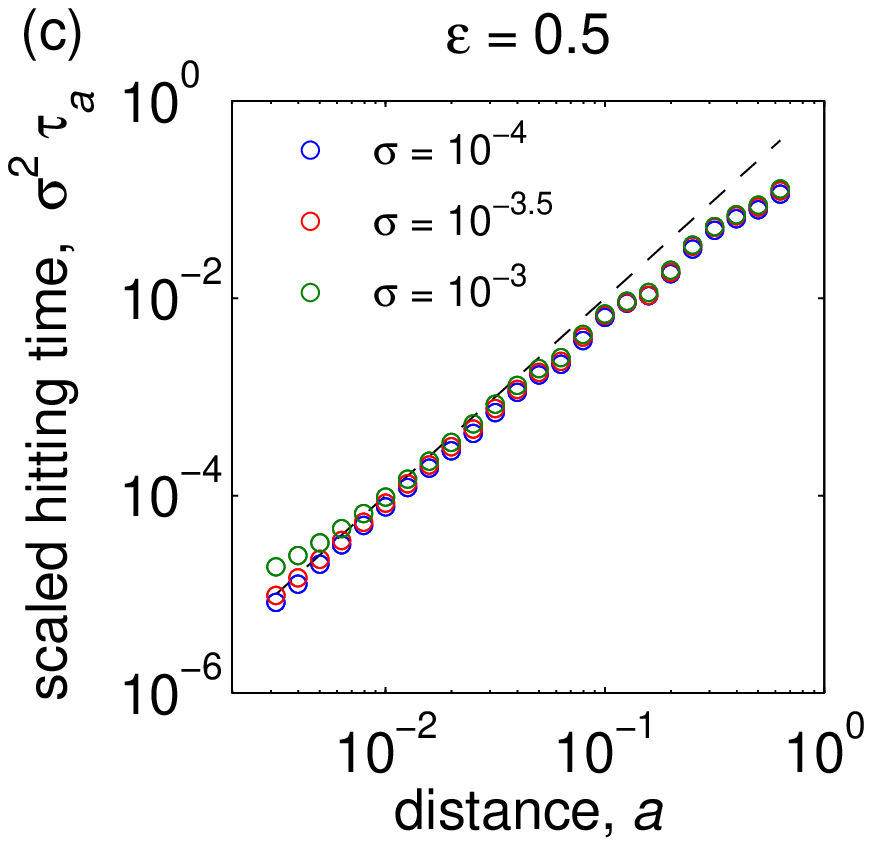, clip =,width=0.32\linewidth } 
\caption{Scaled hitting times. For nonlinearity parameters $\epsilon=0.01$ (a), $0.2$ (b), and $0.5$ (c), we plot the scaled mean hitting time $\sigma^2 \tau_a$ vs distance $a$ for noise intensities $\sigma = 10^{-4}$, $10^{-3.5}$, and $10^{-3}$, which are plotting in blue, red, and green circles, respectively. Each data point represents the mean over $10^4$ realizations. The Brownian motion approximation $\sigma^2\tau_a=a^2$ is plotted as a black dashed curve.} \label{fig4}
\end{figure*}

We begin by noting that the dependence of the hitting time on noise level can be scaled out of the Brownian motion approximation in Eq.~(\ref{eq:Brownian}) by considering the rescaling $\tau_a\mapsto\sigma^2\tau_a$. Therefore, we plot in Fig.~\ref{fig4} the scaled quantities $\sigma^2\tau_a$ vs the distance $a$ using the same results presented in Fig.~\ref{fig3}. Results using $\epsilon=0.01$, $0.2$, and $0.5$ are plotted in panels (a)--(c), respectively, and results using $\sigma=10^{-4}$, $10^{-3.5}$, and $10^{-3}$ are plotted in blue, red, and green circles, respectively. As expected, for small $\epsilon$, e.g., panel (a), the results collapse onto the curve $\sigma^2\tau_a=a^2$. More surprisingly, however, we observe that the results for larger $\epsilon$, i.e., panels (b) and (c), also collapse on one another, even in the regime where the Brownian motion approximation fails.

The fact that even the deviations from the Brownian motion approximation scale is remarkable and suggests two important points. First, the observed speed-up in transport is not solely a result of the added stochasticity, but also from the underlying nonlinear dynamics which are deterministic. Second, the transport we observe falls into two regimes. We refer to these two regimes as a {\it linear} transport regime and a {\it nonlinear} transport regime. The linear transport regime corresponds to dynamics that adhere well to the Brownian motion approximation, indicating that the dynamics are driven primarily by the linear stochastic term and subsequently the expected hitting time is well-described by the power-law $\tau_a\propto a^2$. The nonlinear transport regime corresponds to dynamics that fail to be described well by the Brownian motion approximation and where the underlying nonlinear dynamics of the standard map contribute to the significant speed-up in transport.

Finally, these results beg for the answer to the question of what the underlying cause for the transition from linear to nonlinear transport. While we find that this effect depends significantly on the structure of the dynamics, and thus on both the nonlinearity parameter and the initial conditions, we find that transport is linear in regions of phase space far from low-order resonance which are dominated by rotational circles. Conversely, transport is nonlinear in resonant regions of the phase space which are dominated by secondary circles. This can be understood as follows. In regions of phase space primarily populated by rotational circles the dynamics push the trajectories primarily in the angle direction, so that the majority of the motion in the action direction is driven by the stochasticity and is captured well by the Brownian motion approximation. On the other hand, in regions of phase space primarily populated by secondary circles, the underlying dynamics can induce a significant displacement in the action variable in just a few iterations. Thus, in these regions transport is facilitated by the nonlinear dynamics which can transport much quicker than the stochasticity can.

As an example, consider the case of $\epsilon=0.5$, the hitting times and scaled hitting times for which are presented in Figs.~\ref{fig3}(c) and \ref{fig4}(c). We observe that the deviation from the Brownian motion approximation, i.e., the kink in the results, occurs roughly at $a\approx0.05$. In Fig.~\ref{fig2} we can see that this is approximately the distance from the golden mean circle to the resonance centered around $y=0.5$. Another sizable kink occurs at $a\approx0.25$, which corresponds to the distance from the golden mean circle to the resonance centered around $y=1$. 

\section{Nonlinear Transport: Another Example}\label{Sec:5}
\begin{figure*}[ht]
\centering
\epsfig{file =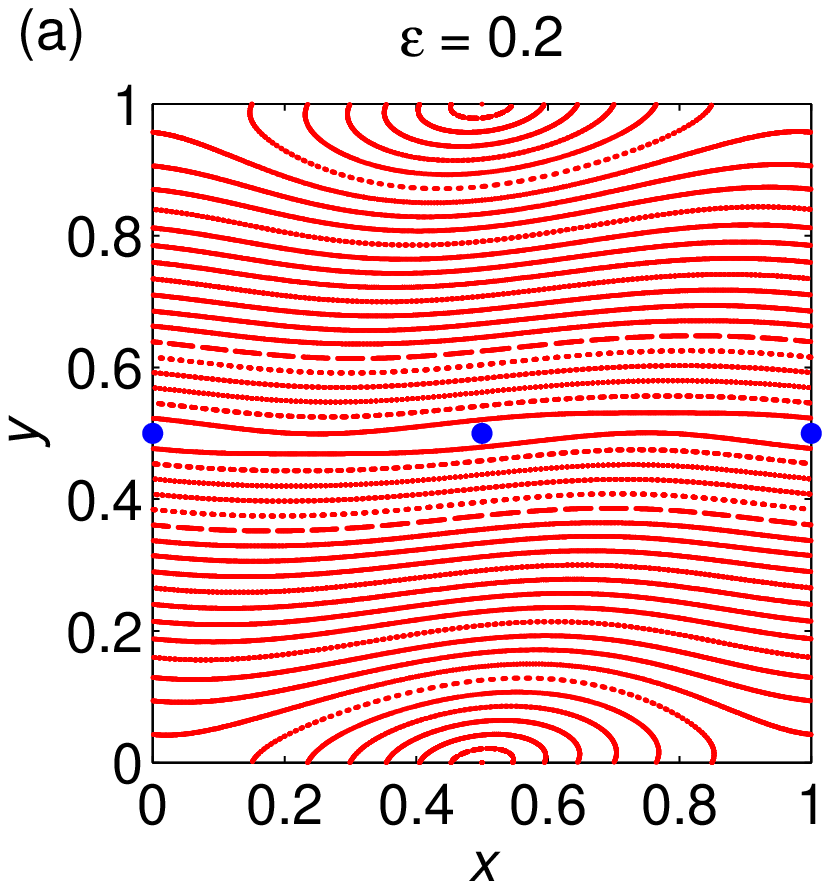, clip =,width=0.32\linewidth }
\epsfig{file =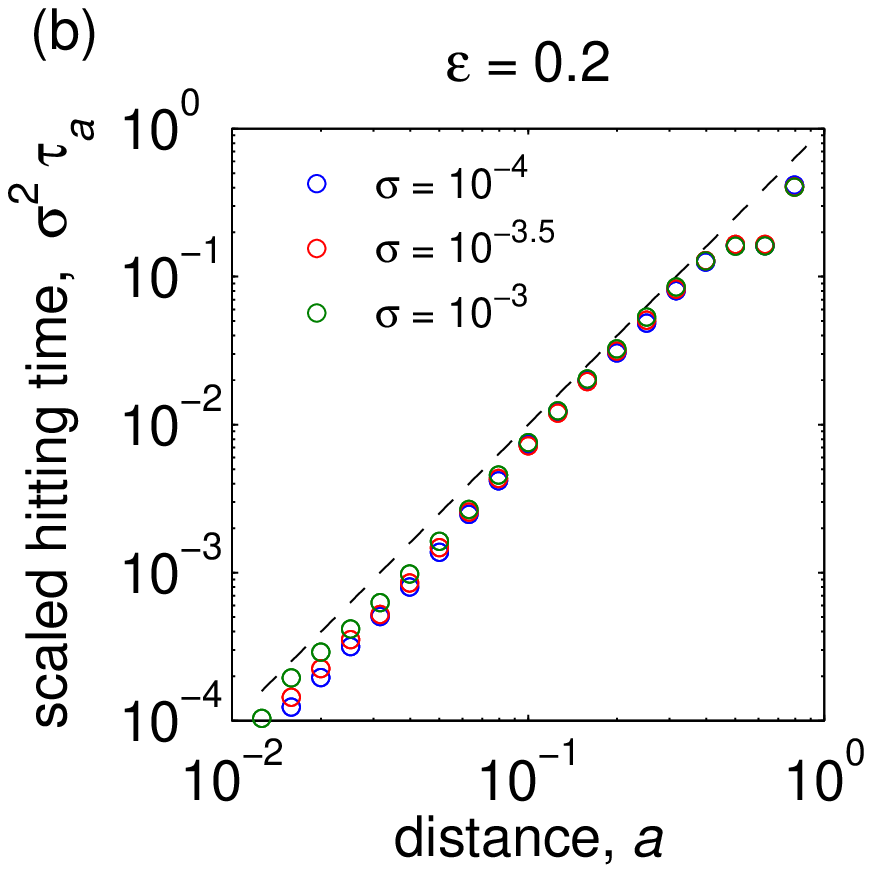, clip =,width=0.32\linewidth } 
\epsfig{file =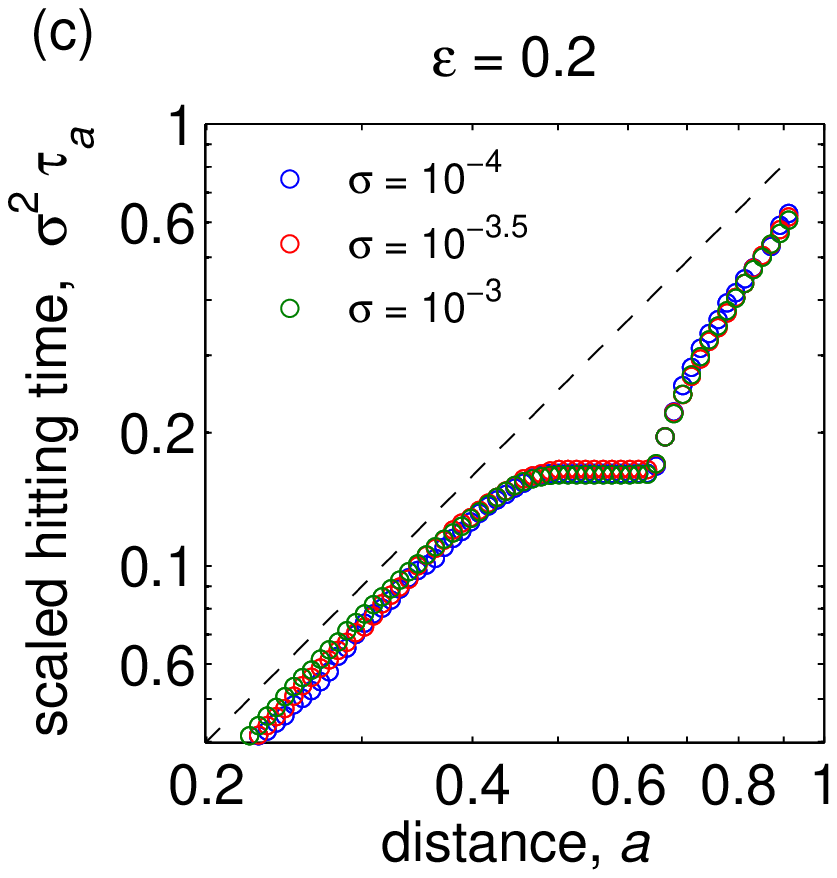, clip =,width=0.32\linewidth } 
\caption{Example two. (a) The orbit $\Phi_0=\{(0,\frac{1}{2}),(\frac{1}{2},\frac{1}{2})\}$ plotted in blue dots, compared to other trajectories in red for $\epsilon=0.2$. (b) Scaled hitting times $\sigma^2\tau_a$ vs distance $a$ for noise intensities $\sigma=10^{-4}$, $10^{-3.5}$, and $10^{-3}$, plotted in blue, red, and green circles, respectively. Each data point represents the mean over $10^4$ realizations. The Brownian motion approximation $\sigma^2\tau_a=a^2$ is plotted as a dashed black curve. (c) A zoomed-in view on the flat region in panel (b) with finer resolution.} \label{fig5}
\end{figure*}

To further demonstrate the nonlinear transport effects in the dynamics of Eqs.~(\ref{eq:SSM}) we consider a modified scenario where, rather than starting at the golden mean circle or another rotational circle, we begin at the period-two orbit $\Phi_0=\left \{ (0,\frac{1}{2}),(\frac{1}{2},\frac{1}{2}) \right \}$. This orbit is plotted as blue dots in Figure~\ref{fig5}(a), with other trajectories plotted in red, for $\epsilon=0.2$. Conveniently, this orbit is fixed for every value of $\epsilon$, making it an illustrative choice of initial conditions for the following experiment.

We proceed by simulating the dynamics of Equations~(\ref{eq:SSM}), again calculating the mean hitting times $\tau_a$ as a function of the distance $a$. We note that, since the orbit is not a graph over the angle variable, we calculate distance simply as the $y$-displacement from the value $y_0=0.5$. In Figure~\ref{fig5}(b) we plot the mean hitting times scaled by the noise squared, $\sigma^2\tau_a$ vs the distance $a$ for $\sigma=10^{-4}$ (blue circles), $10^{-3.5}$ (red circles), and $10^{-3}$ (green circles), again averaged over $10^4$ realizations. The Brownian motion approximation $\sigma^2\tau_a=a^2$ is denoted by the dashed black curve. Note that the results for different values of $\sigma$ collapse nicely when scaled appropriately. 

In Figure~\ref{fig5}(b) we observe two nonlinear transport effects, manifesting as significant deviations from the linear transport along the Brownian motion approximation. First, and most subtle, we observe that for small distances (i.e., $a\lesssim10^{-1}$) there is a speed-up in the hitting times. Upon further investigation, we find that this turns out to be due to the curvature of the trajectories around the period-two orbit depicted in Figure~\ref{fig5}(a). In particular, the mildly sinuous shape of the orbits - an effect of the nonlinear dynamics of the system - causes a small speed-up in transport with respect to the action variable. Second, we see a more pronounced nonlinear transport effect at larger distances (i.e., the second- and third-to-last data points) where the hitting times flatten out. Since the logarithmic scaling of the plot may diminish the apparent size of this region, we plot in Figure~\ref{fig5}(c) a zoomed-in view with a ten-fold finer discretization. Here we see more clearly the dramatic, flat section of the plot in the range $0.4\lesssim a\lesssim0.6$ where transport occurs very quickly. Looking back to the foliation of the phase-space in Figure~\ref{fig5}(a), we note that this flat region corresponds almost precisely with the distances from the initial value $y_0=\frac{1}{2}$ to and extending through the large resonances centered at $(\frac{1}{2},0)$ and $(\frac{1}{2},1)$. Once past these large resonant regions the transport returns to the linear regime, as shown by the final point in Figure~\ref{fig5}(b).

\section{Discussion}\label{Sec:6}
In this paper we have studied the dynamics of Chirikov's standard map with an added stochasticity term [see Eqs.~(\ref{eq:SSM})]. The added noise term in the stochastic standard map destroys the invariant manifolds present in the noiseless case that organize the phase space and bound transport. The destruction of these invariant objects facilitates widespread, unbounded transport in the action direction not present in the noiseless case. Using appropriately defined hitting times, we have quantified the transport that occurs in the stochastic standard map and found that transport falls into two broad categories: \emph{linear} transport and \emph{nonlinear} transport. In the case of linear transport, movement in the action direction is dominated by the stochastic term and is well-described by a simple Brownian motion such that hitting times scale with the square of the distance. In the case of nonlinear transport, the stochasticity combines with the underlying nonlinear dynamics of the map to facilitate a significant speed-up in the mean hitting times. Importantly, we find that linear transport prevails in regions of phase-space dominated by non-resonant dynamics, while nonlinear transport prevails in regions of phase-space dominated by resonant dynamics - which become more pronounced as the nonlinearity parameter of the dynamics is increased.

The effect of added stochasticity to the standard map or other conservative systems has been studied in a handful of other works, e.g., see Refs.~\cite{Froeschle1975ASS,Karney1982PhysD}, however very few recent results exist. To our knowledge, this is the first study concerned with the effect that added stochasticity has on the the transport that occurs as a result of breaking the invariant objects in phase space. Subsequently, our results open new questions for further investigation. First, the examples presented in this paper were chosen using parameter values such that the noise-less phase space was primarily foliated by circles and (periodic) fixed points. One interesting avenue for investigation will be to study how transport occurs as larger chaotic seas emerge (i.e., at larger values of the nonlinearity parameter). We hypothesize that the emergence of significant chaotic seas will cause even more significant speed-up in transport than do resonant circles. Second, we have used as a primary example in this paper Chirikov's standard map due to its simplicity and widespread popularity. However, the effect that noise has on other, possibly more complicated conservative systems - discrete or continuous - will be an interesting question to ask. Finally, we find it remarkable that, although the invariant manifolds organizing the phase space in the noiseless case are broken with the addition of noise, the dynamics remain remarkably robust. An investigation into the limits of this robustness, e.g., in terms of the noise intensity and/or the nonlinearity parameter, would be worthwhile.

\bibliographystyle{plain}

\end{document}